\documentclass[aps,preprint,epsfig,showpacs]{revtex4}
\usepackage{graphicx}
\usepackage{amssymb}
\usepackage{epsfig}
\begin{document}
\title{Statistical mechanics of  coil-hairpin transition 
in a single stranded DNA oligomer}

\author{Sanjay Kumar, Debaprasad Giri$^\dagger$ and Yashwant Singh}

\affiliation{Department of Physics, Banaras Hindu University,
Varanasi 221 005, India, \\
$^\dagger$Physics Section, MMV, Banaras Hindu University,
Varanasi 221 005, India.}


\begin{abstract}
A model of self-avoiding walk with suitable constraints on
self-attraction is developed to describe the conformational
behavior of a short RNA or a single stranded DNA molecule that 
forms hairpin structure and calculate the properties associated 
with coil-hairpin transition by enumerating all possible 
conformations of a chain of N monomers in two and three dimensions. 
The first and last five monomers of the chain have been allowed 
to pair and form the stem of the hairpin structure while the 
remaining monomers can form a loop.  The coil-hairpin transition 
is found to be first order with large entropy change. While 
the rate of unzipping of the hairpin stem is found to be independent 
of the length of the loop and the dimensionality of the space, 
the rate of closing varies greatly with loop length and dimensionality 
of the space.
\end{abstract}

\maketitle

The hairpin structure is often observed in RNA and single 
stranded DNA (ssDNA) molecules and is known to participate 
in biological functions such as  the regulation of gene 
expression, DNA recombination, and facilitation of mutagenic 
events etc \cite{1,2,3,4,5}.  The stability and conformational 
fluctuations of the hairpin structure have recently been 
investigated by designing simple ssDNA oligonucleotides \cite{6,7,8,9,10}
that have few complimentary bases at both ends of the chain and one 
type of nucleotides in the middle, {\it e.g.} $5'-${\bf CCCAA} - 
$(X)_m$ - {\bf TTGGG}$-3'$ where $X$ is any one of the four 
nucleotides and $m$ its number. A hairpin structure has two 
structurally and dynamically distinct domains: a base paired 
stem and a single stranded loop connecting two halves 
of the stem. The stem shows the same response to change in solution 
conditions as a dsDNA oligomer. The loop region, however, shows a 
wide range of folding patterns that depend on the number  and type 
of bases  in the loop.  The stem-loop structure fluctuates thermodynamically
between different conformations, which in a simplified description, 
are divided into two main states: the open state when all the 
binding monomers are separated and the fully closed one where all
the complimentary bases are paired. However, as shown below, this 
simplification is not always correct.

The closed-to-open transition requires the large energy to unzip the 
base pairs of the stem whereas 
the closing transition requires the two arms of the loop come close to 
each other in space in such a way that hydrogen bonding interaction 
between the complimentary nucleotides can take place. The understanding 
of the nature of the open-to-closed transition is essential to our 
understanding of biopolymer dynamics. 

By attaching donor and accepter flurophores to both ends of a ssDNA the 
open-to-closed conformational dynamics of the hairpin have been investigated 
\cite{6,7,8,9}. The fraction of the open state is displayed in terms of the 
melting curve, which depicts the variation of the static fluorescence 
intensity with temperature. The melting temperature $T_m$ of the 
structure is defined as the temperature where the probabilities of the 
closed and open states are equal. In such experiments it is, however, 
difficult to tell whether the closed state corresponds to a hairpin 
structure with some base pairs intact [Fig. 1(a-c)] or whether it is 
merely a folded state where the donor and accepter are quite close 
without base pairing [Fig. 1(d-f)]. 

\begin{figure}
\includegraphics[width=2.8in]{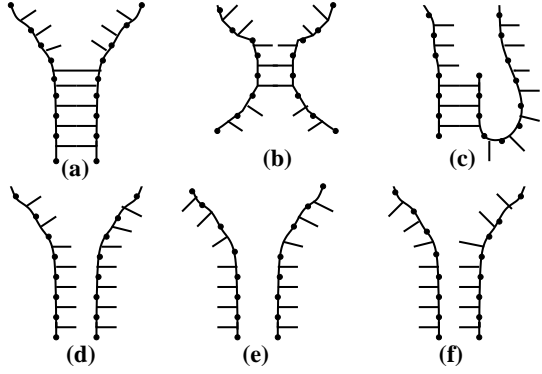}
\caption{Schematic representation of possible configurations 
of bases of the two ends of a chain on reaching close to 
each other. In (a) all the complementary bases are paired 
whereas in (b) and (c) the pairing is only partial. In situations
shown in (d) to (f) pairing cannot take place even though two 
ends of the chain are close to each other. }
\end{figure}

Attempts have recently been made to understand the conformational 
fluctuations of hairpin-loop structure in telehelic chain using 
Monte Carlo simulations \cite{11,12}. In these calculations 
one or two attractive sites with spherically 
symmetric interactions have been attached at both ends of a chain of 
hard spheres. However, as stated above, in ssDNA and RNA the hairpin 
structure is formed due to pairing of complementary bases through 
hydrogen bonds which are highly directional and therefore, spherical 
nature of attraction  taken in these studies does not represent the 
real situations. 

In this communication we propose a lattice model of self-avoiding 
walk with constraints that mimic the condition of pairing of bases 
and calculate the conformational behaviour of the hairpin-loop 
structure. A self-avoiding walk of N steps (N+1 vertices) is considered 
on a square lattice in two-dimensions (2D) and on a cubic lattice in 
three dimensions (3D). A step of the walk represents a monomer 
of the chain. With each monomer a base is attached which has a direction 
associated with it.  The first $n$ bases of the walk represent nucleotide,
say A (or C) and the last $n$ bases the complementary nucleotide 
T (or G). The remaining $N-2n=m$ bases represent one type of  
nucleotides which do not participate in pairing. The repulsion 
between monomers at short distances ({\it i.e} excluded volume) 
is taken into account by the condition of self-avoidance. 

The pairing between bases can take place  
only when the bases at the two ends of the chain representing 
the complementary nucleotides approach on the neighbouring lattice 
bonds with their directions pointing to each other (see Fig. 1(a)).  
A base can at most pair with a 
complementary base.  The base pairing can not take place when 
the two bases representing complementary nucleotides approach on 
neighbouring lattice bonds but with their directions not pointing 
to each other as shown in Fig. 1(d-f).  In Fig 1(b) and (c) a 
situation in which only partial pairing can take place is shown. 

All possible conformations of ssDNA of $N$ bases mapped by the 
self-avoiding walks \cite{13}  with the constraints specified above 
and having steps $N \le 27$ 
on a square lattice (in 2D) and $N \le 19$ on a cubic lattice 
(in 3D) have been exactly enumerated \cite{14,15}. 
The partition function of the system is found from the relation, 
\begin{equation}
Z_N (T) = \sum_{i=0}^{n} C_N(i) (e^{-\epsilon/k_B T})^i
\end{equation}
where $C_N (i)$ is the total number of configurations corresponding to
walk of $N$ steps with $i$ number of intact base pairs. In all the 
results reported below we have taken $n=5$ and $\epsilon = -0.08$ eV.

The partition function defined by Eq. (1) has six terms corresponding 
to six values of $i$ from 0 to 5 ($i = 0$ corresponds to the open 
state, $1 \le i \le 4$ to the partially bound state and $i=5$ to the 
fully bound state of the hairpin-loop structure). We can calculate 
the probability of these states for a chain of length $N$ from the 
relation 
\begin{equation}
P_i (T) = \frac{Z_N^i (T)}{Z_N (T)}
\end{equation}
where $Z_N^i(T) = C_N(i) e^{-i \epsilon/k_B T}$ and $ Z_N (T) = 
\sum_{i=0}^{5} Z_N^i$.

Let the probability of the closed state which includes both partially 
and fully closed configurations of the chain is defined as 
\begin{equation}
P_c (T) = \frac{Z_N^c(T)}{Z_N(T)}
\end{equation}
where $Z_N^c (T) = \sum_{i=1}^{5} Z_N^i$.

The values of $P_i(T)$ of partially bound states ({\it i.e.} for 
$1 \le i \le 4$) are found to be small compared to the values of 
$P_0(T)$ and $P_5(T)$ but not always negligible. Our results 
show that as the length of the part of the chain that forms loop 
in hairpin-loop structure increases, the values of $P_i(T)$ of the 
partially bound states decreases and is expected to be negligibly 
small when $m >> 2n$. However, when $m \lesssim 2n$, the contribution 
due to partially bound states are significant and cannot be neglected.
The values of $P_i(T)$ is found to peak at 
some temperature and the position of the peak shifts to the lower 
temperatures as $i$ is increased from 1 to 4. The peak height decreases 
on increasing the chain length. In 2D for chains of small length, 
$N \le 21$, the peak of $P_3 (T)$ is found higher than that of $P_4(T)$ (see 
Fig. 2), but 
for $N \ge 23$ the peak of $P_4(T)$ becomes higher than that of  $P_3(T)$. 
In 3D we, however, find that the peak of $P_4(T)$ is higher than that of 
$P_3(T)$ 
for all chain lengths studied by us. In Fig. 2, we plot the values of $P_i(T)$  
found in 2D and 3D for a chain having $N=19$ monomers to show the temperature 
dependence and magnitude of $P_i(T)$ of partially bound states. 
We note that both the peak height and the width of $P_i(T)$ are larger 
in 2D than in 3D. This shows that the probability of finding partially
bound states is less in 3D than in 2D. This is because of the competition 
between thermal fluctuations which forces the chain to be in the open state 
and pairing of bases which take place when  complimentary monomers 
approach each other in a particular way. Since the contribution arising 
due to thermal fluctuations in 3D is higher and the energy gained 
due to formation of  pairs is same in 3D and 2D, the partially bound 
states are less stable in 3D than 2D. 

\begin{figure}
\includegraphics[width=4in]{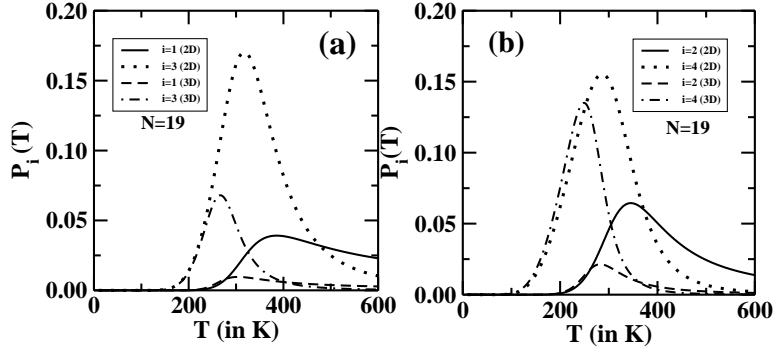}
\caption{Comparison of probability ($P_i(T)$ of partially 
closed states ($1 \leq i \leq 4$) for a chain of  $N=19$ monomers 
in two and three dimensions. (a) $i=1, 3$ and (b) $i=2, 4$. }
\end{figure}

\begin{figure}
\vspace {.5in}
\includegraphics[width=4in]{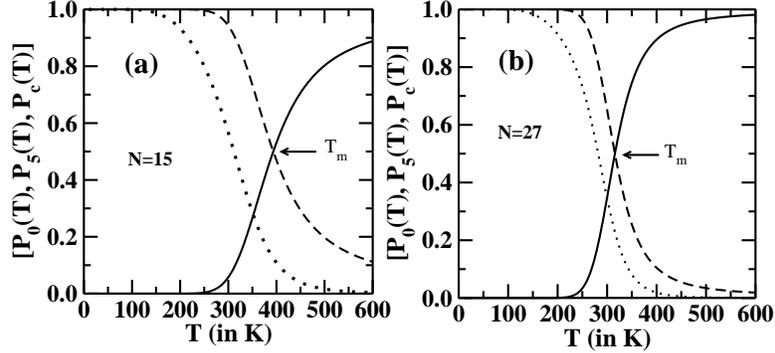}
\caption{Comparison of probability $P_i(T)$ of opened $(i=0)$ [solid line], 
completely closed $(i=5)$ [dotted line], and sum of partially and  
fully closed states $(\sum_{i=1}^5 i)$ [dashed line]  as a function  
of temperature $T$, in two dimensions. (a) N=15,  (b) N=27.}
\end{figure}

The values of $P_0(T)$, $P_5(T)$ and $P_c(T)$ as a function of $T$ 
are plotted in Fig. 3 for chains of $N=15$ and $27$ in 2D and in Fig. 4 
for $N=15$ and $19$ in 3D. The difference in the values of $P_c(T)$ 
and $P_5(T)$ is due to partially bound states which are included 
in $P_c(T)$ but not in $P_5(T)$. The difference, as expected, seems 
to decrease on increasing the length of the chain and in going from 
2D to 3D. The melting temperature $T_m$ ({\it i.e.} the temperature 
at which transition from the hairpin-loop state to coil state takes 
place) is found from the intersect of $P_0(T)$ and $P_c(T)$ which 
takes place at their values of 0.5. We list in Table I the values 
of $T_m$ for different chain lengths in both 2D and 3D cases. The 
value of $T_m$ is found to have strong dependence on the chain 
length ($T_m$ decreases as chain length increases) and on the 
dimensionality of the space.

The Helmholtz free energy and other thermodynamic quantities like the 
internal energy ($U$), the entropy ($S$) and the specific heat ($C$) can 
be found from the standard relations;
\begin{eqnarray}
F &=& -k_B T \ln Z_N (T) = -k_B T(\ln Z_N^0 - \ln P_0(T)) \\
U &=& \sum_{i=1}^{5} i \epsilon  P_i (T) \\
S &=& - \left ( \frac{\partial F}{\partial T} \right ) = k_B \ln Z_N + 
\frac{U}{T} \\
C &=& - T \left ( \frac{\partial^2 F}{\partial T^2}\right )
\end{eqnarray}
In general,the  transition is located from the maxima of $C(T)$ curve. 
The temperature $T_c$ at which maximum of specific heat is found, is lower 
than $T_m$. Using Eqs. (4) and (7) and the fact that $Z_N^0$ is 
temperature independent and $P_0 (T_m) = \frac{1}{2}$ one can calculate 
the difference $(T_m -T_c)$ for a given chain. We list in Table I the 
value of $T_c$ for several values of $N$ and note that as $N$ increases 
$T_c$ approaches to $T_m$ and the peak in $C(T)$ curve becomes sharper. 
The calculated entropy shows a jump from low value corresponding 
to the hairpin structure to a value equal to $N \ln \mu$ ($\mu$ being the 
connectivity of the lattice for a SAW) at high temperatures. 

\begin{table}
\caption{Values of transition temperature and change in entropy at the 
transition temperature $T_m$.} 
\vspace {.1in}
\begin{tabular}{|llllllllll|lllllllll|}\hline
 & & & & & 2D & & & & & & & & & & 3D& & & \\
\hline
$N$ && && $T_c$ && $T_m$ && $\frac{\Delta S}{k_B}$ && $N$ && && $T_c$ &&
$T_m$ && $\frac{\Delta S}{k_B}$  \\
&& && (in K) && (in K) &&  && && && (in K) && (in K) 
 && \\
\hline
15 && && 346 && 393 && 3.82 && 13 &&  && 337 && 359 && 5.02  \\
19 && && 323 && 352 && 4.71 && 15 &&  && 300 && 311 && 6.32  \\
23 && && 310 && 330 && 5.33 && 17 &&  && 282 && 392 && 6.93  \\
27 && && 300 && 315 && 5.91 && 19 &&  && 271 && 279 && 7.37  \\
\hline
\end{tabular}
\end{table}

\begin{figure}
\vspace {.5in}
\includegraphics[width=4in]{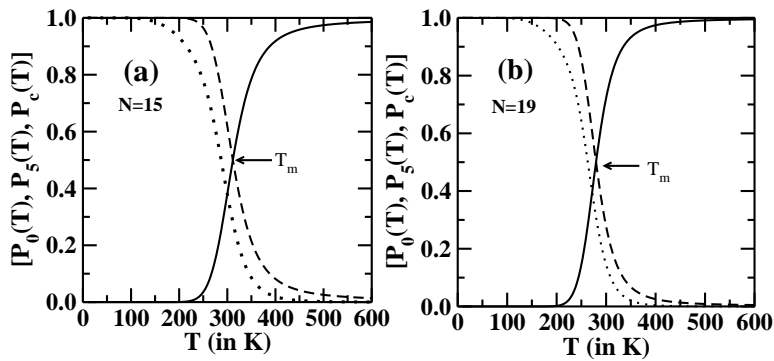}
\caption{Same as in Fig. 3, but in three dimensions and for (a) N=15 and 
(b) N=19. }
\end{figure}

The jump in entropy at $T_m$ is found from the relation 
\begin{equation}
\frac{\Delta S}{k_B} = - \frac{U (T_m)}{k_B T_m}
\end{equation}

We list the value of $\Delta S$ in Table I. The large change in 
entropy indicates that the transition is first order. 

From the principle of detailed balance one gets 
\begin{equation}
\frac{k_{o-c}}{k_{c-o}}= \frac{P_c}{P_0}
\end{equation}
where $k_{i-j}$ is the rate coefficient jumping from $i$ to $j$ state. 
The rate coefficient is assumed to follow the Arrhenius kinetics, 
$k_{i-j} = k_{i-j}^{*} \exp(-\beta \Delta F_{i-j})$,
where $\Delta F_{i-j}$ denotes the free energy barrier associated 
with jumping from $i$ to $j$ states and $k_{i-j}^{*}$ is a 
constant depending on the chemical nature of the chain. 
When the conformation changes  from the open state to the closed 
one, the free energy barrier $\Delta F_{o-c}$ corresponds only to the 
entropy loss from a random coil to a ring polymer; and hence it should 
be temperature independent for a given chain. On the other hand, when 
the structure fluctuates from the closed state to the open one, the 
energy barrier is the binding energy {\it i.e.} $\Delta F_{c-o} \simeq 
U$. 

\begin{figure}
\includegraphics[width=4in]{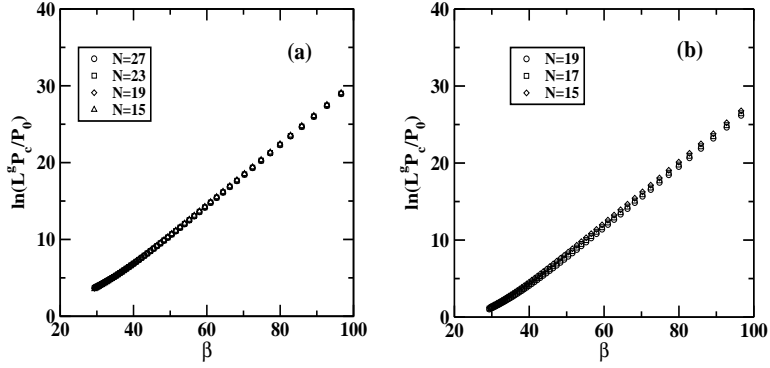}
\caption{$\ln(\frac{L^{g} P_c}{P_0})$ is plotted against $\beta$ 
for several chain lengths N; (a) in two dimensions with $L=m+1$ and 
$g=1.82$ and (b) in three dimensions with $L=m+1$ and $g=2.15$. 
Slope in both cases is equal to the internal energy of the fully 
bound state.}
\end{figure}

The probability that a polymer of length $L$ folds on to itself to 
form a ring is given by 
\begin{equation}
p_L= \frac{\sum_{i=1}^{5} C_L (i)}{\sum_{i=0}^{5} C_L (i)}
\end{equation}
We find the main contribution to $p_L$ is due to the term 
corresponding to formation of one pair. As suggested above 
$k_{o-c}$ is proportional to $p_L$. For a self-avoiding walk of 
length $L$ one finds \cite{16} 
\begin{equation}
p_L \propto L^{-g}
\end{equation}
where $g = (d\nu + \gamma - 1)$.
Here $d$, $\gamma$ and $\nu$ are, respectively, the dimensionality 
of space, susceptibility and correlation length critical exponents. 
Substituting the value of $\gamma$ and $\nu$ one finds $g = 1.83$ in 
2D and $1.97$ in 3D. From our results we find that $g = 1.82$ in 
2D and $g = 2.15 $ in 3D and $L = m+1$. When we compare these 
values of $g$ with those found from Eq. (11) a very good agreement is seen 
in 2D, but in 3D the value found by us is higher. However, the value of 
$g$ found by Monte Carlo simulation on 3D lattice is 2.15 \cite{17,18}
and the analytical estimates place its value in the range 
$2.10 \leq g \leq 2.20$ \cite{19}. 

Using Eq. (9) we write 
\begin{equation}
\frac{1}{k_{c-o}}= \frac{1}{k_{o-c}} \frac{P_c}{P_0} = \frac{1}{L^g k_{o-c}}
\frac{L^g P_c}{P_0}
\end{equation}
We plot in Fig. 5 $\ln(\frac{L^g P_c}{P_0})$ with $L = m+1$ and $g = 1.82$ 
in 2D and 2.15 in 3D as a function of $\beta$ for several 
chain lengths in 2D and 3D. 
As can be seen from these figures that the values collapse into a single  
line with a slope equal to $0.4\pm0.002$. The slope is same in both 2D and 3D 
cases and equal to the value of the internal energy of the fully closed 
state. This result indicates that regardless of the loop length and 
dimensionality of the space the thermal fluctuations provide a probability 
of $\exp(-\beta U)$ to unbind the closed conformation.

We have presented a of self-avoiding walk on a lattice with 
on-site repulsion and appropriate constraints on self-attraction 
to describe the conformational behaviour of a single stranded DNA and
RNA molecules that form hairpin structure. For a given chain 
of $N$ monomers, we have enumerated all possible configuration of 
the chain and using this data we calculated properties associated with 
the coil-hairpin transition. The transition is found to be first 
order with large entropy change. The transition temperature ($T_m$) 
is found from the intersection of the probability corresponding to
the open state and the closed state which includes both partially 
and fully closed states. These probabilities are found to intersect 
at their values of 0.5. The transition temperature ($T_m$) decreases
on increasing the loop length as has been observed in a real system 
\cite{6}. The transition temperature, $T_c$ found from the maximum 
of specific heat curve has the value lower than $T_m$ and the gap between 
the two decreases as the length of the chain increases. The probability 
of the partially closed states are small and decreases on increasing 
the part of the chain that forms loop. The rate of unzipping of the 
hairpin stem was found to be proportional to $\exp(-\beta U)$ and 
independent of both the length of the loop and dimensionality of the 
space. The rate of closing is, however, found to depend rather 
strongly on both the loop length and the dimensionality of the
space. 

\acknowledgments
We thank Navin Singh for many helpful discussions. 
Financial assistance from $INSA$, New Delhi and DST, New Delhi
are acknowledged.


\begin{thebibliography}{0}
\bibitem{1} Zazopoulos, E.; Lalli, E.; Stocco, D. M.; Sassone-Corsi, P.  
{\it Nature} (London) {\bf 1997}, 390, 311.
\bibitem{2} Froelich-Ammon, S. J.; Gale, K. C.; and Osheroff, N. {\it 
J. Biol. Chem.} {\bf 1994}, 269, 7719. 
\bibitem{3} Trinch, T. Q.; Sinden, R. R. {\it Genetics} {\bf 1993}, 134, 
409.
\bibitem{4} Tang, J.; Temsamani, J.; Agarwal, S. {\it Nucleic Acids Res.} 
{\bf 1993}, 21, 2729.
\bibitem{5} Tyagi, S.; Kramer, F. R. {\it Nature Biotechnol.} {\bf 1996}, 14, 
303. Liu, X.; Tan, W. {\it Anal. Chem.} {\bf 1999}, 71, 5054.
\bibitem{6} Bonnet, G.; Krichevsky, O. and Libchaber, A. {\it Proc. Natl. 
Acad. Sci. U.S.A.} {\bf 1998}, 95, 8602.
\bibitem{7} Goddard, N. L.; Bonnet, G.; Krichevsky, O.; Libchaber, A.
{\it Phys. Rev. Lett.} {\bf 2000}, 85 2400.
\bibitem{8} Wallce, M. I.; Ying, L.; Balasubramanian, S. and Klenerman, D. 
{\it J. Phys. Chem. B} {\bf 2000}, 104, 11551.
\bibitem{9} Ying, L.; Wallce, M. I.; Klenerman, D. {\it Chem. Phys. Lett.}
{\bf 2001}, 334, 145.
\bibitem{10} Vallone, P. M.; Paner, T. M.; Hilario, J.; Lane, M. J.; 
Faldasz, B. D.; Benight, A. S. {\it Biopolymers} {\bf 1999}, 50, 425.
\bibitem{11}  Sheng, Y.-J.; Chen, J. Z. Y.; Tsao, H.-K. 
{\it Macromolecules} {\bf 2002}, 35, 9624.
\bibitem{12} Tsao, H. K.; Chen, J. Z. Y.;  Sheng, Y. J. 
{\it Macromolecules} {\bf 2003}, 36, 5863, Sheng, Y. J.; Lin, H. J.; 
Chen, J. Z. Y.; Tsao, H. K.  {\it J. Chem. Phys.} {\bf 2003}, 118, 8513.
\bibitem{13} Foster, D. P. and Seno, F. {\it J. Phys. A: Math. Gen.}
{\bf 2001}, 34, 9939.
\bibitem{14} Vanderzande, C. {\it Lattice models of polymers}; Cambridge
University Press: UK, 1998.
\bibitem{15} Singh, Y.; Giri, D.; Kumar, S.
{\it J. Phys. {\bf A}:Math and Gen.} {\bf 2001}, 34, L67.
\bibitem{16} De Gennes, P. -G. {\it Scaling Concepts in Polymer Physics} 
Cornell Univ. Press: Ithaca, 1993. Des Cloiseaux, J.;  
Jannink, G. {\it Polymers in Solution}; Clarendon, Oxford University 
Press, 1990.
\bibitem{17} Carlon, E.; Orlandini, E.; Stella, A. L. {\it Phys. Rev. 
Lett.} {\bf 2002}, 88, 198101. 
\bibitem{18} Baiesi, M.; Carlon, E.; Stella, A. L. {\it Phys. Rev. E}
{\bf 2002}, 66, 021804. 
\bibitem {19} Kafri, Y.; Mukamel, D.; Peliti, L. {\it Phys. Rev. Lett.} 
{\bf 2000}, 85, 4988.
\end{thebibliography}
\end{document}